\documentclass[twocolumn]{aastex631}
\usepackage{graphicx}
\usepackage{booktabs}
\usepackage{booktabs}
\usepackage{multirow}
\usepackage{gensymb}
\usepackage{float}
\usepackage{xcolor}
\usepackage{array}
\usepackage{tabularx}
\usepackage{amsmath}

\newcommand{\Teff}{\ensuremath{T_\mathrm{eff}}}
\newcommand{\logg}{\ensuremath{\log g}}
\newcommand{\Msun}{\mathrm{M}_\odot}
\newcommand{\pc}{\mathrm{pc}}

\begin{document}

\title{Dynamical Origins of Azimuthal Metallicity Variations in the Galactic Disk: \\Insights from Kinematic Ridges with Gaia}

\correspondingauthor{Carlos Jurado}
\email{carlosjurado@utexas.edu}
\author[0009-0009-7568-8851]{Carlos Jurado}
\affiliation{Department of Astronomy, The University of Texas at Austin, 2515 Speedway Boulevard, Austin, TX 78712, USA }

\author[0000-0002-1423-2174]{Keith Hawkins}
\affil{Department of Astronomy, The University of Texas at Austin, 2515 Speedway Boulevard, Austin, TX 78712, USA }

\author[0000-0001-8917-1532]{Jason A. S. Hunt}
\affiliation{School of Mathematics \& Physics, University of Surrey, Guildford GU2 7XH, UK}

\author[0000-0002-3855-3060]{Zoe Hackshaw}
\affil{Department of Astronomy, The University of Texas at Austin, 2515 Speedway Boulevard, Austin, TX 78712, USA }

\author[0000-0001-5522-5029]{Carrie Filion}
\affiliation{ Center for Computational Astrophysics, Flatiron Institute, 162 Fifth Avenue, New York, NY 10010, USA}

\author[0000-0002-6411-8695]{Neige Frankel}
\affiliation{Canadian Institute for Theoretical Astrophysics, University of Toronto, 60 St. George Street, Toronto, ON M5S 3H8, Canada}

\author[0000-0002-5840-0424]{Christopher Carr}
\affiliation{Department of Astrophysical Sciences, Princeton University, Princeton, NJ 08544, USA}

\begin{abstract}
Kinematic and spectroscopic studies in the past few years have revealed coherent azimuthal metallicity variations across the Milky Way’s disk that may be the result of dynamical process associated with non-axisymmetric features of the Galaxy. At the same time, stellar kinematics from \textit{Gaia} have uncovered  ridge-like features in the velocity space, raising the question of whether these chemical and dynamical substructures share a common origin. Using a sample of disk stars from \textit{Gaia} DR3, we find that azimuthal metallicity variations are correlated with kinematic ridges in the $V_\phi$–$R$ plane, suggesting a shared origin. We utilize a suite of Milky Way test-particle simulations to assess the role of transient spiral arms, the bar, and interactions with a Sagittarius-like dwarf galaxy in simultaneously shaping both chemical and kinematic substructures. Among the physical mechanisms explored, bar and spiral arm interactions are the ones that consistently reproduce both the chemo-kinematic features and alignment observed in the \textit{Gaia} data. While our model of an interaction with a Sagittarius-like dwarf galaxy can also induce kinematic and metallicity substructure, the amplitude of the azimuthal metallicity variations are too weak, suggesting this is likely not the dominant influence. Although additional contributing processes cannot be ruled out, the azimuthal metallicity variations observed in \textit{Gaia} are best explained by a dynamical origin. Our results support the view that that azimuthal metallicity variations in the Galaxy are driven by similar dynamical mechanisms responsible for generating the kinematic ridges and co-moving groups.

\end{abstract}

\section{Introduction}
\label{sec:intro}
The European Space Agency's \textit{Gaia} mission \citep{Gaia2016+} has been revolutionary for our understanding of the Milky Way (MW). \textit{Gaia} Data Release 3 \citep[DR3;][]{GaiaDR320222+} has provided us with 5D astrometry measurements (Parallax, Right Ascension, Declination, Proper motion in Right Ascension, and Proper motion in Declination) for more than a billion stars, radial velocity measurements of 33 million stars, and fundamental astrophysical parameters for over 5 million stars from the onboard Radial Velocity Spectrometer \citep{Recio-Blanco2023+}. This has enabled us to study the chemo-dynamical structure and evolution of the Galaxy with unprecedented detail.

Prior to \textit{Gaia}, early studies of stellar kinematics were already beginning to recognize substructures around the solar neighborhood, including classical co-moving groups \citep{Eggen1958a, Eggen1958b}. These are collections of stars around the solar neighborhood that share similar velocities and were historically interpreted as remnants of disrupted clusters or star-forming regions. However, the Hipparcos mission \citep{Perryman1997+} marked a turning point in our understanding of the co-moving groups and stellar kinematics by providing parallaxes and proper motions for roughly $\sim100,000$ stars near the solar neighborhood \citep{Skuljan1999}. With this improvement, large-scale kinematic and spectroscopic studies began to challenge the idea that the co-moving groups originated from disrupted clusters. The lack of chemical homogeneity among stars in these groups and the large number of stars within these groups supported a dynamical origin instead \citep{Dehnen1998, Bovy2010+, Bensby2014+}.  

In the era of \textit{Gaia}, it has now become possible to map out the kinematic structure of the Galactic disk far beyond the solar neighborhood for millions of stars \citep{Hunt2025+}. The precision and shear number of stellar parallaxes and proper motions measured by \textit{Gaia} have revealed a rich network of disequilibrium structures including ridges and arches in the velocity distribution of disk stars \citep[e.g.][]{Kawata2018+, Fragkoudi2019+, Khanna2019+, GaiaCollaboration2018}, extended radial structures in the action-angle phase space \citep[e.g.][]{Trick2019+}, and spiral shells in the vertical velocity space \citep[e.g.][]{Antoja2018+}. We now know that the co-moving groups are local manifestations of these broader features seen in the velocity distribution of disk stars \citep[e.g.][]{Trick2019+, Fragkoudi2019+, Hunt2019+}. 

The various kinematic substructures can originate from non-axisymmetric potentials, such as those that arise from the Galactic bar \citep[e.g.][]{Dehnen2000, Fux2001+, Perez-Villegas2017+}, spiral arms \citep[e.g.][]{Quillen2011+,Hunt2018+,Hattori2019+,Khoperskov2022+}, or even external gravitational interactions \citep[e.g.][]{Purcell2011+, Laporte2018+, Laporte2019+, Huntb2021, Gandhi2022+}. Disentangling the root dynamical mechanism and the corresponding stellar response is challenging, but significant progress has been made in doing so.  It has been shown that the co-moving groups can be created from resonances with specific combinations of transient spiral arms and/or the bar \citep{Hunt2018+, Hunt2019+, Hattori2019+}. Together,  phase-mixing and resonances with spiral arms and the bar can naturally create many of the the kinematics ridges and arches seen in the stellar velocity distribution \citep{Hunt2018+, Huntb2018+, Martinez-Medina2019+}.

While stellar kinematics offer clues about past dynamical interactions, a more complete picture emerges when combining stellar motions with their chemical composition, setting the stage for a deeper understanding of the MW’s structure. The chemical composition of stars provides another powerful tool for unraveling the history of the MW. Stellar elemental abundances encode information about a star's birthplace and evolutionary history \citep{Freeman2002+}, offering a complimentary perspective on the structure and formation of the Galaxy. It has been conclusively shown that the MW's thin disk exhibits a negative radial metallicity profile, where the interior is more metal-rich than the outer regions \citep[e.g.][]{Luck2011+, GaiaDR3Chemistry, Yan2019+}. This negative metallicity profile, in combination with the stellar age distribution, has been taken as evidence of the `inside-out' formation for the MW \citep{Larson1976+, Chiappini1997+, Frankel2019+}. 

More recently, large-scale stellar surveys have suggested the presence of chemical substructure superimposed on the radial metallicity profile, particularly in the form of azimuthal metallicity variations of order $\sim 0.1$~dex \citep{Poggio2022, Hawkins2023, Hackshaw2024}. Potential explanations for these variations include natal origins, such as uneven mixing of the interstellar medium \citep[e.g.][]{Davies2009+, Grand2015+, Spitoni2019+, Sanchez2020+, Khoperskov2023+}, as well as dynamical processes \citep[e.g.][]{DiMatteo2013+,Khoperskov2018+, Wheeler2022+, Filion2023+, Debattista2025+}. \cite{Hackshaw2024} suggests that the observed azimuthal metallicity variations might be dynamical in nature, as they persist even among older stars. Additionally, \citet{Frankel2025+} showed that dynamical perturbations on top of a radial metallicity profile can produce large-scale azimuthal metallicity variations of order $\sim 0.1$~dex. The exact dynamical mechanisms driving these variations can range from interactions with the Galactic bar \citep[e.g.][]{DiMatteo2013+, Filion2023+}, spiral arms \citep[e.g.][]{Grand2016+, Khoperskov2018+, Khoperskov2022+,Debattista2025+}, or even past encounters with the Sagittarius (Sgr) Dwarf galaxy \citep{Carr2022+}. Importantly, many of these dynamical mechanisms are already known to create the co-moving groups and kinematic ridges in the velocity distribution of disk stars. This connection raises the possibility that the azimuthal metallicity variations are not an independent phenomenon, but rather another manifestations of the same underlying dynamical processes.

In this work, we investigate if the known azimuthal metallicity variations in the Galactic thin disk are shaped by similar dynamical mechanisms responsible for producing the kinematic ridges and co-moving groups. To facilitate this investigation our paper is outlined as follows: Section \ref{sec:data} describes the \textit{Gaia} sample used in this analysis and Section \ref{sec:models} presents a suite of MW-like simulations designed to test how the bar, spiral arms, and interactions with a Sgr-like dwarf galaxy shape both kinematic and chemical substructure. Section \ref{sec:method} outlines our methodology for calculating the radial metallicity profile, recovering the observed azimuthal metallicity variations, and our process for identifying the kinematic substructures/co-moving groups. The results in Section \ref{sec:results} and \ref{sec:disc} demonstrate that dynamical processes are a major driver of the observed azimuthal metallicity variations and these chemical variations are  correlated with the kinematic substructures/co-moving groups. We summarize which Galactic features best reproduce the observed chemo-kinematic alignment and provide our concluding remarks in Section \ref{sec:conclude}. 
\section{The Gaia Data}
\label{sec:data}


In order to investigate the origins of stellar azimuthal metallicity variations, we need spatial, kinematic, and chemical information for a large population of thin disk stars. The combination of chemo-dynamical information provided by \textit{Gaia} DR3, along with it's broad coverage over the Galactic disk, makes it an ideal dataset for this work. 

\begin{figure}
    \centering
    \includegraphics[width=1\linewidth]{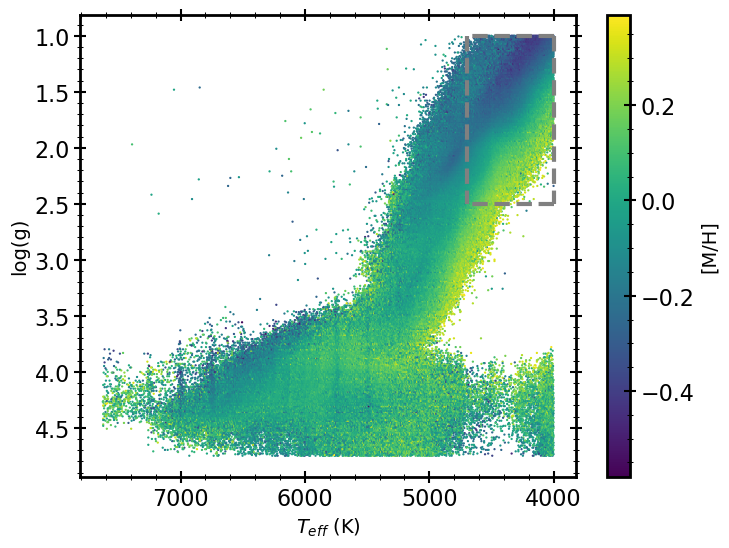}
    \caption{Kiel Diagram of the full Gradient Analysis Sample stars. The grey box represents the subsample of stars that fulfill the \logg{} and \Teff{} constraints in the text of Section \ref{sec:data}. }
    \label{fig:kiel_diagram}
\end{figure}

We draw the initial sample from the Gradient analysis sample located in \citet{GaiaDR3Chemistry} which can be created from the ADQL Query found in Listing 3 in Appendix B of their paper. This sample is designed to provide high-quality metallicity and 6D phase-space coordinates for stars with an effective temperature, \Teff{} $>$ 4000K and with a renormalised unit weighted error, RUWE  $\leq$ 1.4. Figure \ref{fig:kiel_diagram} shows the Kiel diagram for all of the stars in \textit{Gaia}'s Gradient analysis sample which consists of approximately 2.7 million stars. The stellar astrophysical parameters associated with this sample are derived from the General Stellar Parametriser from spectroscopy (GSP-Spec) module, part of \textit{Gaia}’s chemo-physical characterization described in \citet{Blanco2023+}. GSP-Spec estimates stellar parameters such as the effective temperature, \Teff{}, surface gravity, \logg{}, and the mean metallicity, [M/H], based solely on spectra obtained by \textit{Gaia}’s onboard Radial Velocity Spectrometer (RVS), a near-infrared, moderate-resolution spectrograph (R = 11,500) covering the 846–870 nm wavelength range.


We determine the distances to each star by cross-matching with the distance catalog from \citet{Bailer-Jones2021+} and utilize the photogeometric distances. We remove stars from the sample with parallax errors and photogeometric distance errors larger than 30$\%$ to ensure fairly precise spatial information. Our focus is exclusively on a subsample of bright giants with an effective temperature ranging from 4000 K to 4700 K and surface gravities (\logg{}) between 1 and 2.5, as measured by GSP-Spec. By selecting a uniform sample of bright giants, we ensure broad coverage over the Galaxy. After invoking the \Teff{} and \logg{} selection cuts, the sample contains $\simeq 975,000$ stars across a broad range of stellar ages and confined to the region within the dashed gray box of Figure \ref{fig:kiel_diagram}. 

Using Astropy's \texttt{SkyCoord} package \citep{Astropy}, we convert the 6D astrometric data to spatial and velocity values in a left-handed Galactocentric coordinate system. We adopt the solar location at $R_\odot=8.3$~kpc \citep{Gillessen2009+} and $z_\odot=27$~pc \citep{Chen2001+}. The local standard of rest (LSR) velocity is set as $V_{LSR}=220$~km/s \footnote{There exists a spread in estimates for the LSR speed in the literature. We explored the impact of $V_{LSR}$ on our results by conducting a test in which $V_{LSR} = 232.8$~km/s \citep{McMillan2017} and found that our main results presented in Section \ref{sec:results} were not impacted.} \citep{Bovy2015} with the sun's velocity relative to the LSR defined by $\vec{V} = (11.1, 12.24, 7.25 )$~km/s \citep{Schonrich2010+}. Throughout this work, we use $R_\odot$ to denote the solar location adopted for transforming the \textit{Gaia} astrometry, meanwhile $R_0$ and $V_0$ refer to the \textsc{Galpy} unit normalizations of the pre-existing test-particle simulations described in Section \ref{sec:models}. 

To account for measurement uncertainties, we performed Monte Carlo error propagation with 500 iterations per star, sampling each astrometric and radial velocity parameter from a Gaussian centered on its reported value with a standard deviation equal to its uncertainty and ignoring all covariances. The mean and standard deviation of the resulting distributions were then adopted as the Galactic position and velocity coordinates and their associated uncertainties. 

Our initial sample includes stars from all Galactic components, not just the thin disk. For each star, we assign thin disk, thick disk, and halo membership probabilities using the kinematic prescription from \citet{ramirez2013+}. Following their Equation (1), which assumes that each Galactic component has a triaxial Gaussian velocity distribution such that the probability of a star belonging to the thin disk ($P_1$), thick disk ($P_2$), and halo ($P_3$) is given by:
\begin{equation}
    \begin{aligned}
    P_i &= \frac{c_i}{(2\pi)^{3/2} \sigma_{U_i} \sigma_{V_i} \sigma_{W_i}} \\
    &\times \exp\left(-0.5\left[ \frac{U^2}{\sigma_{U_i}^2} + \frac{(V-V_i)^2}{\sigma_{V_i}^2} + \frac{W^2}{\sigma_{W_i}^2}  \right] \right)   
    \end{aligned}
\end{equation}
where $(U, V, W)$ is the star's heliocentric Galactic velocities and $c_i$ is a normalization constant to ensure that $\sum P_i = 1$. The values for the component velocity dispersion $\sigma_{U_i}$, $\sigma_{V_i}$, $\sigma_{W_i}$, and mean velocity $V_i$ are taken from footnote (7) in \citet{ramirez2013+}.

We remove stars from our sample with a halo probability greater than 5$\%$ and those with a vertical distance greater than 0.3 kpc from the mid-plane (i.e. the scale height of the thin disk as determined by \citet{Gilmore1983+, Du2003+, Juric2008+}), leaving us with $\simeq 650,000$ stars. Following \citet{ramirez2013+}, we select thin disk stars as stars with a thin disk probability greater than 50$\%$ leaving us with $\simeq 400,000$ thin disk stars. In the rest of this paper, our analysis is based on this thin disk sample. The median error on \Teff{}, \logg{}, and $[M/H]$ are $28$ K, $0.08$ dex, and $0.03$ dex respectively.

\begin{table}
\centering
\resizebox{\columnwidth}{!}{%
\begin{tabular}{cc}
\toprule
{Quantity} &      Average Uncertainty \\

\midrule
\midrule

Right Ascension, $\alpha$ (deg.) & $0.012$~mas \\
Declination, $\delta$ (deg.) & $0.011$~mas \\
Parallax (mas) & $0.016$~mas \\
Proper motion in RA, $\mu_\alpha$ (mas/yr) & $0.016$~mas/yr \\
Proper motion in Dec, $\mu_\delta$ (mas/yr) & $0.014$~mas/yr \\
Line-of-sight radial velocity (Gaia) (km/s) & $0.368$~km/s \\
\midrule
Galactocentric Radius, $R$ (kpc) &    $0.031$~kpc 
\\

Galactocentric Vertical distance, $Z$ (kpc)  &    $0.003$~kpc
\\

Galactocentric Azimuthal angle, $\phi$ (deg.)  &    $0.284$~deg
\\

Galactocentric Radial Velocity, $V_R$ (km/s) &    $0.635$~km/s
\\

Galactocentric Vertical Velocity, $V_z$ (km/s)  &    $0.387$~km/s
\\

Galactocentric Azimuthal Velocity, $V_\phi$ (km/s) &    $0.646$~km/s

\\
\bottomrule
\end{tabular}
}
\caption{Average uncertainty on the 5D \textit{Gaia} astrometry + line-of sight radial velocities, derived galactocentric coordinates, and derived galactocentric velocities of the full thin disk sample. Uncertainties on the galactocentric coordinates were estimated from 500 Monte Carlo samples per star.}
\label{tab:mc_errors}
\end{table}

\section{The Models}
\label{sec:models}

\begin{table*}
    \centering
    \begin{tabular}{lcccccccc}
        \hline
        Model Name & Model Description & $R_b$ [kpc] & $\Omega_b \; (\times \Omega_0)$  & $\phi_b$ [deg] & $N$ & $\theta_{sp}$ [deg] & $R_{sp}$ \\
        \hline
        \texttt{LSB+Spiral} & \texttt{Long Slow Bar And Transient Spiral Arms}  & $5.0$ & 1.3 & $25$ & 2 & 25 & 0.3 \\

        \texttt{SFB+Spiral} & \texttt{Short Fast Bar And Transient Spiral Arms} & $3.5$ & $1.85$ & $25$ & $2$ & $25$ & 0.3 \\

        \texttt{LSB} & \texttt{Long Slow Bar} & $5.0$ & 1.3 & $25$ & -- & -- & -- \\
        \hline
    \end{tabular}
    \caption{Model parameters for the non-impact MW simulations. Parameters listed include the bar radius $R_b$, the bar pattern speed $\Omega_b$, the angle of the bar with respect to the line joining the Galactic center and the sun $\phi_b$, the number of spiral arms N, the pitch angle $\theta_{sp}$ the radius scale length of the arm $R_{sp}$, and the circular frequency at the solar radius $\Omega_o$.}
    \label{tab:models}
\end{table*}

While the \textit{Gaia} dataset allows us to explore the azimuthal metallicity variations, it does not provide immediate insights into their origin. In this Section, we describe the suite of test-particle simulations that will be utilized to explore the effects that the Galactic bar, spiral arms, and interactions with a satellite galaxy have on the creation of  stellar kinematic and metallicity variations. These Galactic structures can perturb stellar orbits, creating both long-term changes in guiding center radius \citep[Churning;][]{Sellwood2002+} and short-term variations. Additionally, non-axisymmetric features can increase orbital eccentricities without a corresponding change in the angular momentum \citep[Blurring;][]{Sellwood2002+}. In this paper, we use the term radial migration to refer to any change (short or long-term variation) in a star's guiding center radius, $R_G$, regardless of the mechanism responsible (e.g. bar, spiral arms, satellite). When we specifically refer to changes in angular momentum at co-rotation, we will call this churning. Radial migration that acts non-axisymetrically alters stellar orbits in a way that produces local deviations from the underlying radial metallicity profile. These variations would manifest as metallicity variations  in the spatial distribution of stars across the disk.

To explore how these dynamical processes create chemo-kinematic substructure, we turn to MW-like simulations that allow us to isolate the effects of the bar, spiral arms, and satellite impacts. For this work, we use the set of simulations from \citet{Hunt2019+, Gandhi2022+}. These are simulations run using \textsc{Galpy} \citep{Bovy2015}, a Python package capable of simulating stellar orbits within an evolving gravitational potential for the MW. The stars in the MW disk of all our simulations are initially sampled from a quasi-isothermal distribution function \citep{Binney2010} using \texttt{galpy.df.quasiisothermaldf}. The distribution function has an initial scale radius $R_s = R_0/3$, local radial velocity dispersion $\sigma_{v_{R}} = 0.15v_c(R_0)$, and local vertical velocity dispersion $\sigma_{v_{z}} = 0.075v_c(R_0)$, where $R_0 = 8 \text{ kpc}$ and $v_c(R_0) = 220 \text{ km/s}$. To ensure equilibrium is reached we evolve the disk using \texttt{galpy.potential.MWPotential2014} for 7 Gyr.



All of our MW simulations that do not include satellite impacts include a bar because its formation and evolution strongly influence stellar migration \citep{DiMatteo2013+, Minchev2013+, Filion2023+, Baba2025+}. To capture this, we use the time-evolving bar potential implemented in \textsc{Galpy}, which allows us to model the bar’s dynamical growth and its effect on stellar orbits over time. The initial bar potential is implemented with \texttt{galpy.potential.CosmphiDiskPotential} and is a generalization of the \cite{Dehnen2000} potential. While we refer the reader to \citet{Huntb2018+} for the full equations governing the bar potential, here we summarize the key model parameters that define the bar. These include the bar radius $R_b$, the bar pattern speed $\Omega_b$, and the angle of the bar with respect to the line joining the Galactic center and then sun $\phi_b$. The amplitude of the bar potential $A_b(t)$ is grown smoothly such that:
\begin{equation}
A_b(t) =
\begin{cases}
0, & \dfrac{t}{T_b} < t_1, \\[6pt]
A_f \left[ \dfrac{3}{16}\,\xi^5 - \dfrac{5}{8}\,\xi^3 + \dfrac{15}{16}\,\xi + \dfrac{1}{2} \right],
    & t_1 \le \dfrac{t}{T_b} \le t_1 + t_2, \\[6pt]
A_f, & \dfrac{t}{T_b} > t_1 + t_2 .
\end{cases}
\end{equation}
{where $t_1$ is the start of bar growth that is set to half the integration time, $t_2$ is the duration of the bar growth, and $T_b = \frac{2\pi}{\Omega_b}$ is the bar period. $\xi$ and $A_f$ are defined as:
\begin{equation}
    \xi = 2 \frac{t/T_b - t_1}{t_2} - 1
\end{equation}
and 
\begin{equation}
    A_f = \alpha_m \frac{v_0^2}{p}\left( \frac{R_0}{R_b} \right)^p
\end{equation}
where $\alpha_m$ is the dimensionless ratio of forces due to the $\cos(m\phi)$ component of the bar potential and the axisymmetric background potential along the bar's major axis.  This bar growth mechanism ensures a smooth transition from the non-barred to the barred state.

For the set of simulations that also include spiral arms, the arms are modeled as corotating transient features commonly seen in N-body simulations \citep{Grand2012+}. The inclusion of corotating arms will maximize the effects of churning, and hence radial migration, because the arms corotate with the stars at all radii. We use \texttt{galpy.potential.SpiralArmsPotential}, which is based on the sinusoidal potential formulation from \cite{Cox2002}. The key model parameters that define the spiral arms are the number of spiral arms N, the pitch angle $\theta_{sp}$ and the radius scale length of the arm $R_{sp}$. In contrast to classical density wave models, there is no fixed pattern speed because the transient arms corotate with the stars at all radii. To model this corotation and the time evolution of the spiral arms, we wrap the potential using \texttt{galpy.potential.CorotatingRotationWrapperPotential} such that: 
\begin{equation}
    \phi \rightarrow \phi + \frac{V_p(R)}{R} \times (t-t_0) + a_p
\end{equation}
where $V_p(R)$ is the circular velocity curve, $t_0$ is the initial time, and $a_p$ is the position angle at $t_0$.

The spiral arms are grown and disrupted smoothly by modulating the amplitude with a time-dependent Gaussian:
\begin{equation}
    A(t) = \text{exp}\left( - \frac{[t-t_0]^2}{2 \sigma^2} \right)
\end{equation}
such that the arms grow from a negligible strength at early times, reach a maximum at $t_0$, and then decay.
The lifetime of the transient spiral arm potential is controlled by the standard deviation of the Gaussian, $\sigma$. The peak amplitude of the spiral is set as $\pm 0.0136\,\Msun\,\pc^{-3}$, yielding an arm-interarm density contrast of 1.31 relative to the local disk density of $0.1\,\Msun\,\pc^{-3}$ \citep{Holmberg2000+}, which is consistent with measurements made by \citet{Drimmel2001+, Benjamin2005+}. Full details of the transient spiral arm setup can be found in \citet{Huntb2018+, Hunt2019+}. 

Including the effects of spiral arms allows us to explore the combined effect that the bar + spiral arms have on stellar migration and chemical substructure. Table \ref{tab:models} presents a summary of the bar/spiral arm MW models and detailed descriptions of these simulations can be found in \citet{Hunt2018+, Huntb2018+, Hunt2019+}. The models listed below have been shown to reproduce some, but not all, of the observed kinematic substructure in the $v_{\phi} - R$ plane and provide insight into which dynamical components may be responsible for different features. 

\subsection{LSB+Spiral Model}
Some more recent observations have suggested that the Galactic bar is longer than previously thought, with a bar radius around $\sim5$~kpc  \citep{Wegg2015+} and with a pattern speed between $32-35$~km $\text{s}^{-1}$ $\text{kpc}^{-1}$ \citep{Clarke2022+, Dillamore2024+, Zhang2024+}. In \citet{Clarke2022+, Hunt2019+}, this bar is commonly referred to as the ``long-slow bar" model. However, other recent studies also report shorter bar lengths \citep[e.g.][]{Lucey2023+}, highlighting how constraining the parameters of the Galactic bar is challenging. Because the bar length and pattern speed is an active area of research, though with a trend towards the long slow bar model, we decide to incorporate two bar models: a long slow bar and a short fast bar (see Section \ref{subsec:shortfastbar}). The long slow bar can reproduce several kinematic substructures seen in the \textit{Gaia} data, including the Hercules moving group via the corotation resonance \citep{Perez-Villegas2017+} As noted in \citet{Hunt2019+}, the inclusion of transient spiral arms can reproduce the missing substructure that the bar alone does not produce. Motivated by this, we incorporate the long slow bar and transient spiral arms model from \citet{Hunt2019+}, hereafter referred to as the \texttt{LSB+Spiral} model, with $R_b = 5$~kpc, $\Omega_b = 1.3 \times \Omega_0$, $\phi_b=25$ degrees, N=2, $\theta_{sp}=25$, and $R_{sp} = 0.3$.

\subsection{SFB+Spiral Model}
\label{subsec:shortfastbar}
In addition to the long slow bar model, we also consider a bar with a shorter bar length, $R_B = 3.5$~kpc, and faster pattern speed of $\Omega_b = 1.85 \times \Omega_0$. This model is based on previous studies of the Galactic Bar by \citet{Lopez-Corredoira2001+, Picaud2003+, Vislosky2024+} and is also capable of producing many of the co-moving groups and kinematic substructure. As an example, the Hercules co-moving group can be produced from the outer Lindblad resonance (OLR) of a short fast bar \citep{Dehnen2000}. Although more recent observations favor a longer and slower bar, the short fast bar model provides a direct comparison of kinematics signatures against the long slow bar model with transient spiral arms. Thus we incorporate the short fast bar and transient spiral arms model from \citet{Hunt2019+}, hereafter referred to as the \texttt{SFB+Spiral} model, with $R_b = 3.5$~kpc, $\Omega_b = 1.85 \times \Omega_0$, $\phi_b=25$ degrees, N=2, $\theta_{sp}=25$, and $R_{sp} = 0.3$.

\subsection{LSB Model}
To isolate the role of the bar in driving stellar migration and chemical substructure \citep[e.g.][]{Filion2023+}, we have a model that only includes the long slow bar but no transient spiral arms, referred to as the \texttt{LSB} model throughout this paper. The \texttt{LSB} model uses the same bar parameters as in the \texttt{LSB+Spiral} model: $R_b = 5$~kpc, $\Omega_b = 1.3 \times \Omega_0$, and $\phi_b=25$ degrees. This setup allows us to determine the bar's ability to produce resonance-induced kinematic features. Unlike models that include spiral arms, the bar-only model yields kinematic substructure that remains relatively stable over time, provided that the speed of the bar pattern remains constant \citep{Hunt2019+}. 

\subsection{Sgr Multi-impact Model}
\label{subsec:satellitesetup}
It has also been demonstrated that interactions with a satellite galaxy can produce both kinematic substructure and azimuthal metallicity variations \citep[e.g.][]{Laporte2018+, Laporte2019+, Huntb2021, Carr2022+}. To explore these effects we also run a simulation of the interaction of a satellite with the disk of a MW-like galaxy. In this work, we focus solely on a Sgr-like multi-impact satellite aimed at testing whether repeated interactions can produce a chemo-kinematic correlation between kinematic ridges and metallicity variations. We note that other satellites, such as the Large Magellanic Cloud, can also influence the Galactic disk \citep[e.g.][]{Laporte2018+, Stelea2024+} however, a detailed comparison of the impacts of difference satellites is beyond the scope of this paper.

{Our goal here is to capture the qualitative disk response due to a Sgr-like multi-impact satellite, not to reproduce the full complexity of the MW-Sgr interaction. Thus, we adopt the simplistic model from \citet{Gandhi2022+} for a Sgr-like satellite, with fixed mass and no tidal stripping, hereafter referred to as the \texttt{Sgr Multi-impact} model. The satellite is modeled as a Plummer sphere using \texttt{galpy.potential.PlummerPotential}, with a total mass of $M_{\text{sat}} = 2 \times 10^{10} M_{\odot}$ and a scale radius of $0.8$ kpc. The orbit is initialized by backwards integrating the present-day position and velocity of Sgr reported in \citet{Vasiliev2020+} with \texttt{galpy.potential.MWPotential2014} and \texttt{galpy.potential.ChandrasekharDynamicalFrictionForce} to account for dynamical friction \footnote{The backwards integration implementation reconstructs a past orbit that is consistent with the present-day Sgr position and velocity under the adopted dynamical friction prescription.}. We integrate the satellite's orbit backwards for 3 Gyr.

The satellite undergoes the first pericenter passage $\sim3$~Gyr ago and the second pericenter passage $\sim1.5$~Gyr ago from the present-day snapshots (see section \ref{sec:present-day}).  Additionally, this model setup excludes both a bar and transient spiral arms for the MW galaxy, ensuring that the satellite is the sole source for any kinematic and metallicity substructures that appear.  

\subsection{The Present-Day Snapshots}
\label{sec:present-day}
To ensure a fair comparison with the \textit{Gaia} thin disk sample, we focus our analysis on the simulation snapshots that provide the best qualitative match to the present-day Milky Way kinematics, hereafter referred to as the present-day snapshot. For the spiral arm and/or bar simulations, we select the present-day snapshots as those identified by \citet{Hunt2019+} as the best qualitative match to the kinematic ridges in the \textit{Gaia} data (e.g. for the \texttt{LSB+Spiral} model this corresponds to the simulation output at an internal  time of $t=-174$~Myr). For the \texttt{Sgr Multi-impact} model, the final snapshot represents the present-day since the current coordinates of Sgr were used for the backwards integration.

\section{Methodology}
\label{sec:method}
In this Section, we outline the methods used to recover the azimuthal metallicity variations and their connection to the stellar kinematics for both the thin disk sample and our simulations. In Section \ref{subsec: Determining Radial Metallicity Gradient}, we establish a procedure to recover the radial metallicity profile. In Section \ref{subsec: Recovering Azimuthal Metallicity Variations} we identify azimuthal metallicity variations and construct a 2D map of metallicity substructure. To aid our comparison of the metallicity variations with stellar kinematics, we identify known co-moving groups in the velocity space in Section \ref{subsec:comovinggroups}.

\subsection{Determining The Radial Metallicity Profile}
\label{subsec: Determining Radial Metallicity Gradient}
The stellar metallicity distribution of stars in the MW disk can be modeled as a 1-D linear function of radius \citep[e.g.][]{Friel2002+, Magrini2009+, Luck2011+, Hayden2014+, Huang2015+, Akbaba2024, Hawkins2023, Hackshaw2024}. Figure \ref{fig:radialgradient} shows the metallicity distribution of our entire stellar thin disk sample as a function of guiding center radius $R_G$, which represents the radius of a circular orbit with the same angular momentum as the star (see Appendix~\ref{appendix: R vs R_G} for a comparison of the results using galactocentric radius $R$ instead). We calculate the guiding center radius of each star from 
$R_G = \frac{R V_\phi}{V_{\mathrm{circ}}}$, 
where $V_{\mathrm{circ}}$ is the circular velocity at radius $R_G$ computed using the 
\texttt{galpy.potential.MWPotential2014} potential. Overlaid as white circles in Figure \ref{fig:radialgradient} is the median [M/H] within 0.2 kpc radial bins that span $R_G \in$ [5kpc, 11kpc]. We perform a linear regression starting outwards of $5$~kpc to avoid the bar/bulge region \citep{Wegg2015+} and within 11 kpc because the radial metallicity profile begins to flatten beyond 12 kpc \citep{Spina2022+}. The best-fitting linear function is overlaid as a solid black line in Figure \ref{fig:radialgradient}. Both the running median [M/H] and the best-fitting linear function clearly show that the inner disk is more metal-rich than the outer disk, consistent with a negative radial metallicity profile across a broad range of ages.


For the simulations, calculating the present-day radial metallicity profile requires painting on the metallicity information for the stars. Although the present-day metallicity profile of the MW is directly observable, it is generally interpreted as the outcome of long-term chemical and dynamical evolution shaped by an inside-out formation scenario where the inner regions of the thin disk formed and enriched more quickly than the outermost regions from a broad range of stellar populations. Both observations and cosmological simulations demonstrate that radial metallicity profiles are not only common in other galaxies \citep{Sander2012+, Sanchez2016+,Sakhibov2018+} but are also a natural result of galaxy evolution across cosmic time \citep{Bellardini2021+}. This suggests that a radial metallicity profile was created early in the thin disk's formation and that the radial metallicity profile has persisted over several Gyr, although its slope has evolved over cosmic time \citep{Anders2017+, Willett2023+}. Motivated by these arguments, we initialize our simulations with a negative radial metallically profile of the form:  
\begin{equation}
    [M/H] = -0.066R_G + 0.65    
\end{equation}
We do this in order to be consistent with the evolution of the radial metallicity profile determined by \citet{Anders2017+}. 

It is important to note that we are not interested in the exact nature of the radial metallicity profile at earlier times, but rather how a given initial profile evolves over time due to dynamics. Thus, we initialize the simulation with a negative radial metallicity profile that is consistent with observations of red giant stars with ages between 1-4 Gyr but this is not intended to precisely replicate the Galaxy’s past chemical state. Our results do not depend on the specific slope of this initial metallicity profile. Since our simulations only track stellar particles and do not include gas physics or feedback any changes in the metallicity distribution arise solely from dynamical processes. As a result, variations in the initial profile affect only the amplitude of azimuthal metallicity variations. Our focus is on identifying the dynamical origin and spatial structure of these variations rather than reproducing their absolute strength.

We calculate the present-day radial metallicity profile in the simulations using the same procedure applied to the \textit{Gaia} thin disk sample. This is done by computing the running median [M/H] of stars in the present-day snapshot within 0.2 kpc bins over $R_G \in [5,11]$ kpc and performing a linear regression. With the calculated radial metallicity profiles, we are now in a position to recover the azimuthal metallicity variations.

\begin{figure}
    \centering
    \includegraphics[width=1\linewidth]{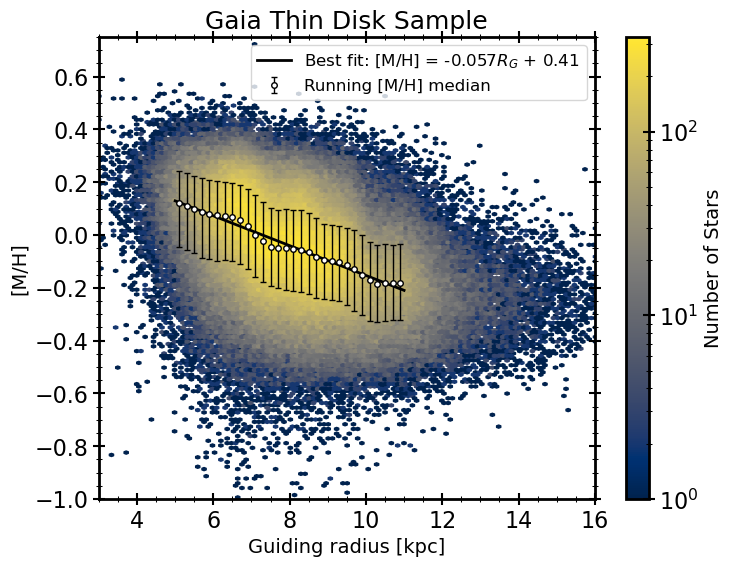}
    \caption{[M/H] distribution of our stellar thin disk sample as a function of the star's guiding center radius. The white circles represent the median [M/H] of all stars in $0.2$~kpc bins with the black vertical bars spanning the 16th-84th percentile range. The best-fitting linear function is overlaid as a solid black line from 5kpc to 11kpc.}
    \label{fig:radialgradient}
\end{figure}

\subsection{Recovering Azimuthal Metallicity Variations}
\label{subsec: Recovering Azimuthal Metallicity Variations}
 Using our modeled radial metallicity profile, we can construct an expected 2D metallicity distribution in the galactocentric X-Y plane that has no azimuthal dependence (See Figure \ref{fig:three_panel}, Middle Panel). The metallicity excess, $\delta [M/H]_{R_G}$, is defined by taking the difference between a star's  `true' metallicity, as measured by GSP-Spec, and the predicted metallicity from the best-fitting radial metallicity model. For the simulations, we compute $\delta [M/H]_{R_G}$ using the star's assigned metallicity from the initial profile as the `true' value and subtracting the predicted metallicity from the present-day radial metallicity model. \footnote{We restrict our analysis of the azimuthal metallicity variations in the simulations to star within a 3.5 kpc radius around the sun to approximate the spatial extent of the thin disk sample. }

\begin{figure*}
    \centering
    \includegraphics[width=1\linewidth]{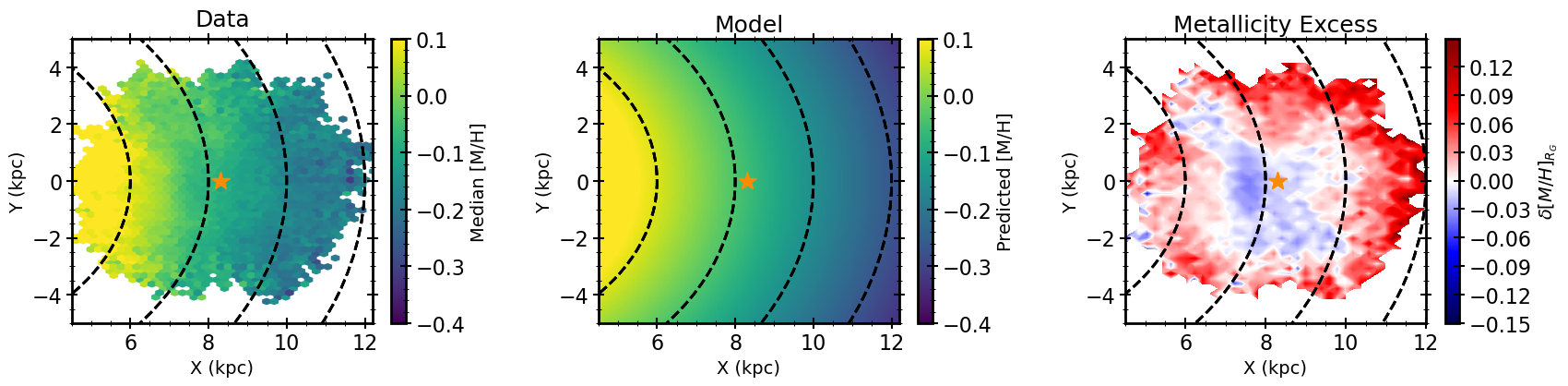}
    \caption{Left Panel: [M/H] distribution of stars in the thin disk sample, plotted in the X-Y galactocentric coordinates. Middle Panel: Best-fitting radial metallicity profile in the X-Y plane. Right Panel: Metallicity Excess ($\delta [M/H]_{R_G}$; Data-Model) in the X-Y plane. The black dashed curves represent circles of radius 6kpc, 8kpc, 10kpc, and 12kpc. The yellow star denotes the sun's location in the X-Y plane. } 
    \label{fig:three_panel}
\end{figure*}

\subsection{Identifying Co-Moving Groups and Kinematic Substructures}
\label{subsec:comovinggroups}

The presence of coherent large-scale azimuthal metallicity variations raises the question of whether dynamical processes are responsible. To address this, we identify the known kinematic substructures and co-moving groups that are dynamically driven and search for a spatial correlation with the metallicity variations.

The upper middle panel of Figure \ref{fig:three_panel_ridges} is the distribution of the thin disk stars in the $V_{\phi}-R$ plane painted by their radial velocities, $V_R$. As noted in the literature \citep[e.g.][]{Antoja2018+, Kawata2018+}, this is a common way to highlight the presence of kinematic substructures across the disk. Distinct inward and outward moving ridges are visible across the velocity distribution, indicating coherent kinematic structures for which some of the prominent co-moving groups are embedded. 

We identify the known co-moving groups by adopting the locations from Figure 12 of \citet{Hunt2019+}. The approximate location of the co-moving groups are labeled in the upper and lower middle panels of Figure \ref{fig:three_panel_ridges} for convenience. Starting in the lower left quadrant, the Hercules streams belong to the two red (outward moving) bands with the uppermost red band corresponding to the main peak of Hercules. The Horn co-moving group is the narrow blue (inward moving) band right above the main peak of Hercules. The red (outwards moving) band above the Horn belongs mostly to the Hyades co-moving group. Above the Hyades is one of the largest substructures, a broad blue band spanning $R = 6$–$10$ kpc and $V_{\phi} = 180$–$280$ km/s, corresponding to the Sirius co-moving group. Additionally, we overlay two dot-dashed lines to trace the slopes of the main peak of Hercules and Sirius, effectively splitting the plot into three regions. The region to the left of the leftmost black dot-dashed line contains the multiple Hercules streams. The middle region, bounded by the two dot-dashed lines, includes the Hyades and Horn co-moving groups, as well as other co-moving groups that are not labeled \citep[e.g][]{Hunt2019+}. The rightmost region lies above the top black dot-dashed line. The $V_{\phi}-R$ plane is split into these three regions solely as an aid to compare with the chemo-kinematic structure seen in the simulations.

With the co-moving groups identified, we are now in a position to present the results of the radial metallicity profiles, metallicity variations, and their relationship to the kinematic space.

\section{Results}
\label{sec:results}

\subsection{The Radial Metallicity Profile and Azimuthal Metallicity Variations}
\label{subsec:results_rmg}

Following our methodology from Section \ref{subsec: Determining Radial Metallicity Gradient}, we find that the best-fitting radial metallicity equation for the thin disk sample as a function of galactocentric radius, R is given by:

\begin{equation}
    [M/H] = -0.062R + 0.45        
\end{equation}

The best-fitting radial metallicity profile for the thin disk sample as a function of the guiding center radius, $R_G$, is:
\begin{equation}
    [M/H] = -0.057R_G + 0.41     
\end{equation}

Our derived $\Delta[M/H]/\Delta R$ falls within the values reported in the literature, despite the age differences of the tracers used, ranging from -0.073 dex $\text{kpc}^{-1}$ for APOGEE giants \citep{Hayden2014+} to -0.045 dex $\text{kpc}^{-1}$ for Cepheids \citep{Lemasle2018+}, and with intermediate values reported from additional studies of open clusters, red clump stars, and other tracers \citep[e.g.][]{Friel2002+, Luck2011+, Onal2016+, Hawkins2023, Hackshaw2024}. Our value for $\Delta[M/H]/\Delta R_G$ is also consistent with other studies focused on different tracers \citep[e.g.][]{Boeche2013+, Plevne2015+, Akbaba2024}. Throughout the remainder of the paper, we focus on the guiding center radial metallicity profile. This choice offers a more physically meaningful framework for connecting stellar chemistry with orbital dynamics \citep{Schonrich2018+, Hunt2020_RG, Khoperskov2022b+, Akbaba2024, Hunt2025+}. Although our conclusions are not sensitive to whether the metallicity profile is calculated using $R$ or $R_G$, we use guiding center radius to provide a clearer interpretation of the observed chemo-dynamical trends.

Using our modeled radial metallicity profile for the thin disk sample, we can search for azimuthal metallicity variations. In the left panel of Figure \ref{fig:three_panel}, we show the median [M/H] distribution of thin disk stars over galactocentric X-Y position. The negative radial metallicity profile is apparent, with stars at smaller galactocentric radii appearing more metal rich relative than those at larger values of R. We also show the best-fitting model for the radial metallicity profile over the X-Y plane in the middle panel. The metallicity excess (defined in Section \ref{subsec: Recovering Azimuthal Metallicity Variations}) over the X-Y plane is shown in the right panel of Figure \ref{fig:three_panel} and reveals clear evidence of azimuthal substructure in the metallicity distribution. The red regions indicate areas that are more metal-rich than predicted by the 1D model, while the blue regions correspond to areas that are more metal-poor. The location and shapes of the metallicity substructures are consistent with the results of \citet{Hawkins2023, Poggio2022} who also utilized thin disk stars from \textit{Gaia}. The strength of the metallicity deviations are on the order of $\sim$ 0.1 dex, consistent with findings from previous studies \citep{Poggio2022, Hawkins2023, Hackshaw2024}.

\begin{figure}
    \centering
    \includegraphics[width=1\linewidth]{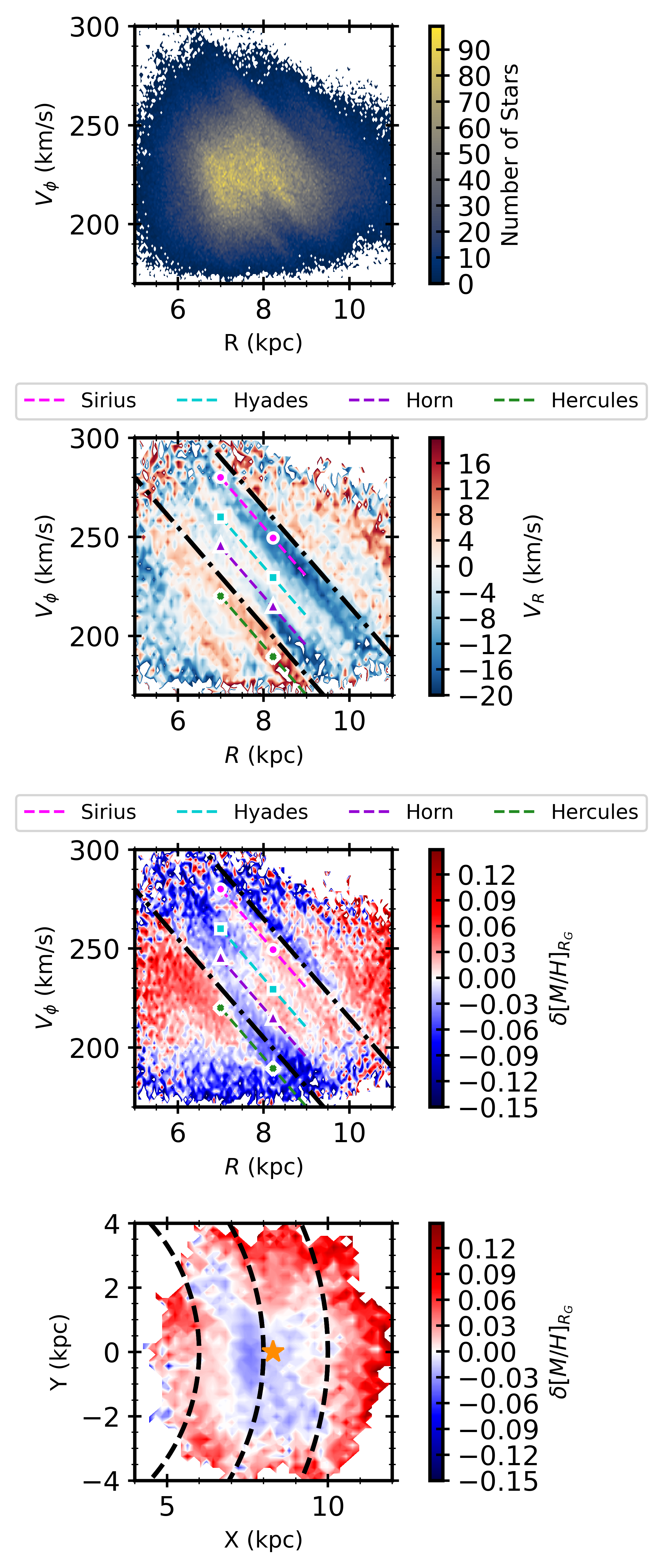}
    \caption{\textbf{Top Panel:} Distribution of azimuthal velocity ($V_{\phi}$) as a function of galactocentric radius (R), colored by number density. \textbf{Upper Middle Panel:} Same as top panel but colored by radial velocity. Overlaid in this panel are the approximate locations of the local co-moving groups. The two black dot-dash lines indicate the slopes of the kinematic ridges associated with the Hercules and Sirius co-moving groups.    
    \textbf{Lower Middle Panel:} Same as top panel but colored by $\delta [M/H]_{R_G}$. 
    \textbf{Bottom Panel:} $\delta [M/H]_{R_G}$ in the X-Y plane. The black dashed curves represent circles of radius 6kpc, 8kpc, and 10kpc. The yellow star denotes the sun's location in the XY plane.}
    \label{fig:three_panel_ridges}
\end{figure}

\subsection{Connecting Azimuthal Metallicity Variations To Kinematic Ridges}

In the preceding section, we established the presence of azimuthal metallicity variations in the thin disk sample. In this section, we examine whether these chemical substructures are correlated with the kinematic ridges in the $V_\phi - R$ plane. From top to bottom, the first three panels of Figure \ref{fig:three_panel_ridges} shows the $V_\phi - R$  plane colored by number density, $V_R$, and $\delta [M/H]_{R_G}$, respectively. The top panel shows clear overdensities of stars that appear as diagonal ridges across a broad range of R values. These ridge-like features can be further enhanced by coloring the $V_\phi - R$ plane by galactocentric radial velocity \citep{Hunt2019+} as is done in the upper middle panel of Figure \ref{fig:three_panel_ridges}. This panel reveals the presence of large-scale kinematic substructures and embedded co-moving groups. The kinematic substructure seen in the $V_\phi - R$ plane and embedded co-moving groups are the result of dynamical interactions with the Galactic bar and spiral arms \citep[e.g][]{Dehnen2000, Quillen2011+, Perez-Villegas2017+, Martinez-Medina2019+} and/or interactions with Sgr \citep[e.g.][]{Khanna2019+, Laporte2019+, Antoja2022+}.

In the presence of a radial metallicity profile, stellar migration will also generate metallicity variations. To explore this, we color the $V_\phi - R$ plane by metallicity excess in the lower middle panel of Figure \ref{fig:three_panel_ridges} to determine if the ridges, which arise from dynamical processes, are accompanied with corresponding chemical signatures. Here, there are also distinct ridges of metal poor and metal rich substructures. To aid the comparison with the upper middle panel, this plot can be divided into three regions, separated by the two dot-dashed black slope lines that follow the main peak of the Hercules and Sirius ridges, respectively. The leftmost region features a metal-rich structure centered around ($\sim$ 6kpc, $\sim230\text{km/s}$). The middle region contains two extended metal-rich bands, while the last region has a single metal-rich band. Metal-poor bands separate each of these metal-rich structures. 

The upper middle panel of Figure \ref{fig:R_ridges_and_variations} clears shows that the metallicity excess substructures have slopes that are aligned with the kinematic ridges. While the slopes of the $\delta [M/H]_{R_G}$ features are aligned with the kinematic ridges, there are instances where the metallicity trend switches along several of the lines (e.g. going from blue to red or vice versa), so the correspondence is not strictly one-to-one. However, the correspondence is significant enough that we can identify metallicity excess structures aligned with specific co-moving group locations. The top of the left-most metal-rich substructure centered at ($\sim$ 6kpc, $\sim230\text{km/s}$) is aligned with the peak of the Hercules feature. The dot-dashed line tracing Sirius also have an associated metal-rich substructure that permeates over several kpc in the disk. There also appears to be a metal-rich structure along the Hyades co-moving group. 

We also conducted a separate test using the APOGEE-astroNN [Fe/H] from the thin disk sample of \citet{Hackshaw2024}. We verify that the chemo-kinematic correlation persists, although the reduced sample size causes neighboring substructures to appear blurred and partially blended. We therefore omit the results from this test to avoid clutter throughout the paper.

The bottom panel of Figure \ref{fig:three_panel_ridges} shows the metallicity excess in the X-Y plane. Here, we see that the chemical substructure that traces the kinematic ridges in velocity space produce coherent azimuthal metallicity variations in the spatial plane. This raises the question of whether the mechanism responsible for generating the azimuthal metallicity variations is related to the one that produces the kinematic ridges in the $V_\phi$–$R$ plane.

\subsection{Comparing The Kinematic Ridges and Azimuthal Metallicity Variations In Simulations}
\label{subsec:comparesims}
In the \textit{Gaia} data, we find a correlation between the locations of the kinematic ridges and the metallicity variations. Although the alignment does indeed exist, we cannot explore which conditions can create this alignment with the data alone. Thus, the simulations allow us to assess whether the presence or absence of a bar, spiral arms or interactions with a satellite galaxy can produce the alignment between the kinematic ridges and metallicity variations.

\begin{figure*}
    \centering
    \includegraphics[width=1\linewidth]{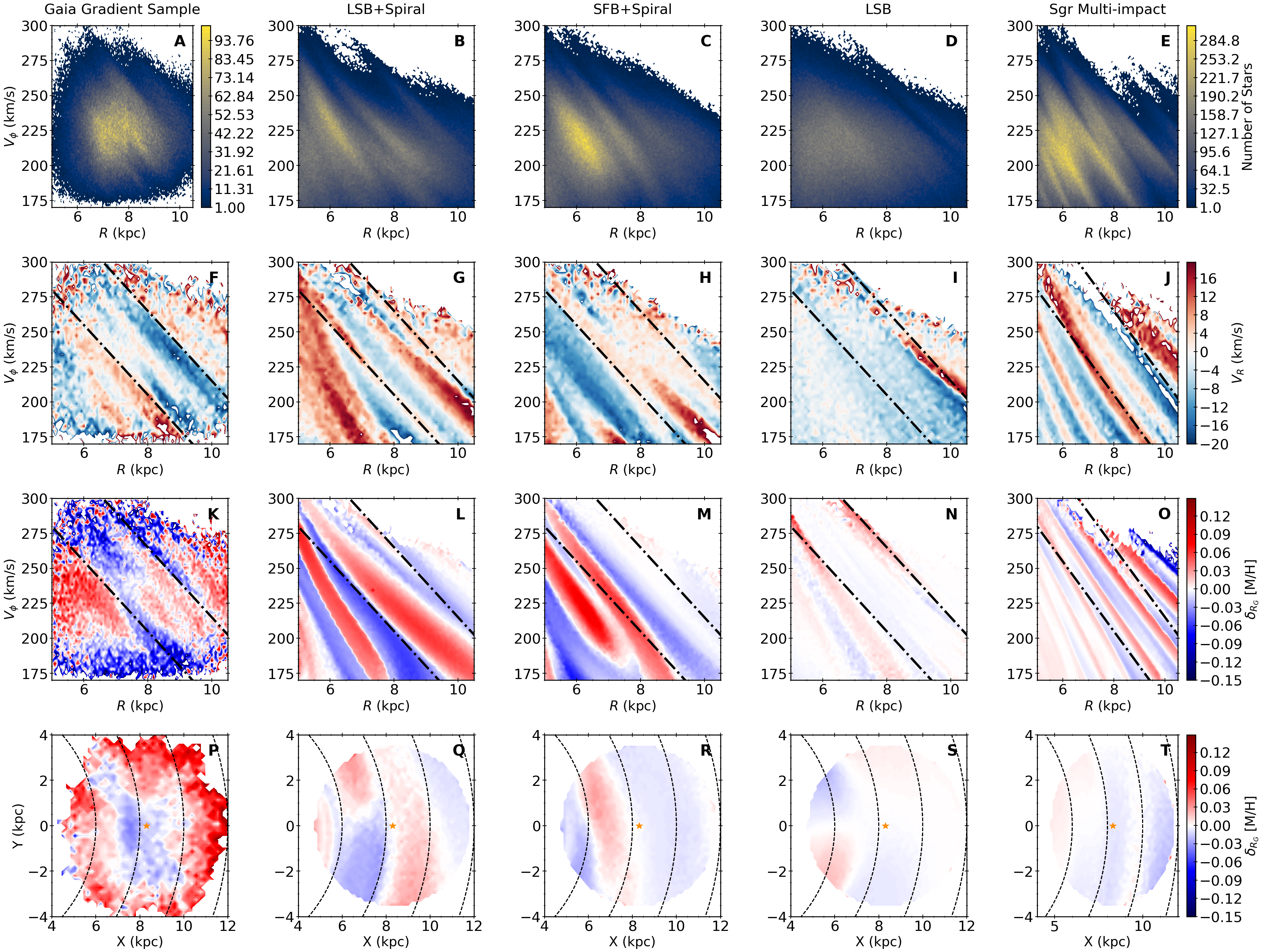}
    \caption{\textbf{First Column:} Distribution of azimuthal velocity ($V_{\phi}$) as a function of galactocentric radius (R), colored by number density. Upper Middle Panel: Same as top panel but colored by galactocentric radial velocity. The two black dot-dashed lines indicate the slopes of the kinematic ridges associated with the Hercules and Sirius co-moving groups.  Lower Middle Panel: Same as top panel but colored by [M/H] excess and overlaid with the dot-dashed lines from the upper middle panel. Bottom Panel: Metallicity Excess in the X-Y plane. The black dashed curves represent circles of radius 6kpc, 8kpc, 10kpc, and 12kpc. The yellow star denotes the sun's location in the XY plane. \textbf{Subsequent Columns:} The same set of plots for the first column but for each set of MW simulations at the present-day snapshot (See Section \ref{sec:present-day} for details on selecting the present-day snapshot). Subplots are labeled A-T solely for convenience when referring to them in the text.}
    \label{fig:megaplot}
\end{figure*}

\begin{figure*}[t]
    \centering
    \includegraphics[width=1\linewidth]{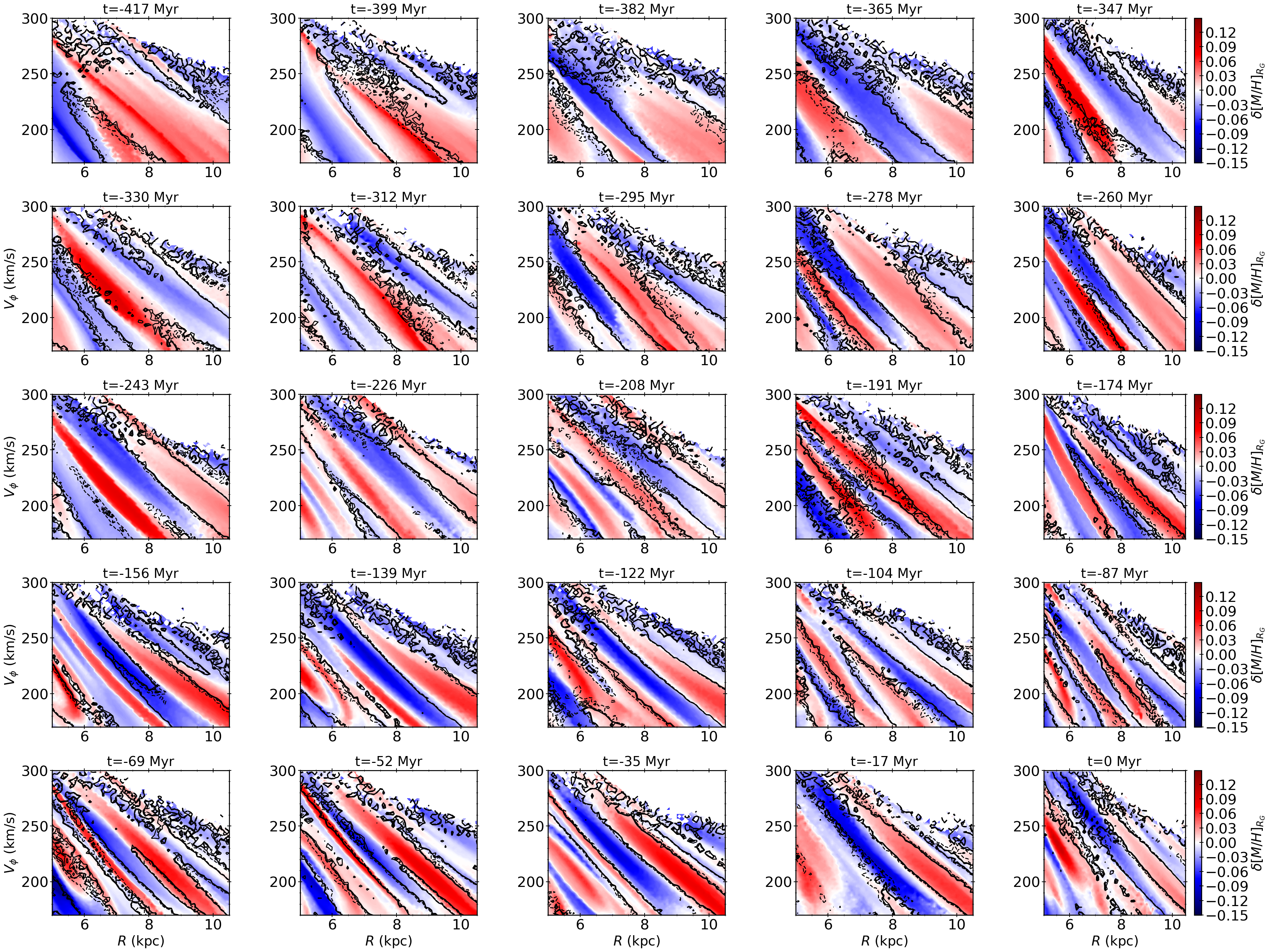}
    \caption{$V_\phi$–$R$ plane colored by guiding center metallicity excess for a series of time steps ranging from t=-417 Myr to t=0 Myr (end of the simulation) for the \texttt{LSB+Spiral} model. The contours in each panel are regions where $V_R=0$ and effectively trace the locations of the kinematic ridges. The present-day snapshot corresponds to the t=-174 Myr timestep.}
    \label{fig:TimeEvolution}
\end{figure*}

\begin{figure}
    \centering
    \includegraphics[width=1\linewidth]{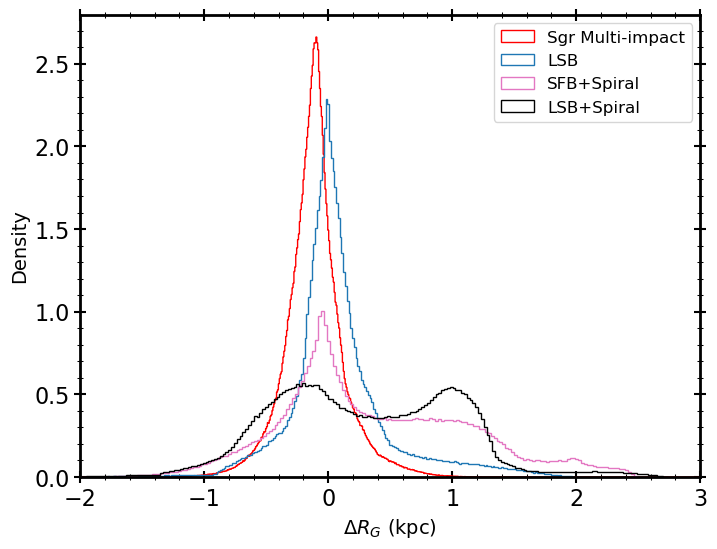}
    \caption{Distribution of $\Delta R_G$ between the various models tested for all stars within 3.5 kpc of the sun at the present-day snapshots. }
    \label{fig:ChangeInRg}
\end{figure}

Figure \ref{fig:megaplot} compares the \textit{Gaia} thin disk sample in the first column with four different MW-like simulations at the present-day snapshots in the subsequent columns. From left to right, the results of the MW-like simulations are shown for the \texttt{LSB+Spiral}, \texttt{SFB+Spiral}, \texttt{LSB}, and \texttt{Sgr Multi-impact} models, respectively. Each column contains four panels that illustrate a complementary view of the chemo-kinematic structure. The top row (Panels A-E) shows the distribution of $V_\phi$ as a function of $R$, colored by number density. The second row (Panels F-J) shows the same plane but colored by $V_R$ to enhance the ridges and co-moving groups in the kinematic space. The black dot-dashed lines overlaid in these panels trace the slopes of the Hercules and Sirius ridges seen in Panel F and are added across the row for comparison. The third row (Panels K-O) illustrates the $V_\phi$–$R$ plane colored by metallicity excess for a direct comparison between the chemical and kinematic substructure. Again, we overlaid the dot-dashed lines from the previous row (Panels F-J). Finally, the bottom row (P-T) projects the metallicity excess onto the galactocentric X-Y plane to demonstrate that metallicity variations in the velocity space manifest as azimuthal metallicity variations in the spatial plane.

In the second column from left, we show the results of the \texttt{LSB+Spiral} model. It's evident that there also exists kinematic ridges and azimuthal metallicity variations \citep{Hunt2019+}. In Panel G, the $V_\phi - R$ colored by $V_R$ plot, there are three main outward-moving (red) bands and 2 inward-moving (blue) bands. The leftmost red band is significantly wider than the other bands in the panel spanning at least 2 kpc across and could be associated with the multiple peaks of a Hercules-like features. The Horn-like, Hyades-like, and Sirius-like features can also be seen in the panel however the exact shapes and strengths of all of these features have slight differences when compared to the data. 

The lower middle panel of the second column (Panel L) displays the metallicity variations in the same plane used to identify the kinematic ridges. Two metallicity excess structures are present, intersecting the $x$-axis at approximately 7–8 kpc and 9.5–10.5 kpc, respectively. Using the same dot-dashed black lines from the last panel and overlaying onto the middle panel shows that the slope of the metallicity excess structure at 9.5–10.5 kpc are aligned with two of the outward co-moving groups in the left panel. At higher $V_\phi$ velocities, there appears to be a tapering of the metallicity substructure that lies between the two dashed lines. The bottom panel (Panel Q) shows the spatial distribution of the metallicity excess structures. Here, we see that the region near the solar neighborhood has a higher than average metallicity excess and demonstrates the existence of azimuthal metallicity variations. In this panel there are two main metallicity excess substructures that occur at an annulus between 4-6kpc and 6-9kpc. The immediate lower left region around the solar neighborhood is metal-poor. Additional hints of metallicity excess substructure are also present near the outer edges of the simulation region.

In the third column from left, we show the results of the \texttt{SFB+Spiral} model. Although kinematic ridges and metallicity variations are still present as in the first two columns, the velocity space is characterized by fewer dominant bands and more localized substructure or patchy substructure. In the $V_\phi - R$ by $V_R$ plot (Panel H), it's unclear whether there are multiple thin bands between the two dashed lines or one dominant outward moving band that has some local substructures of stars that are inward-moving. A well-defined outward-moving ridge is visible, extending from approximately (8 kpc, 220 km/s) to (10 kpc, 170 km/s). In addition, smaller outward-moving clumps are present, centered near (5 kpc, 180 km/s) and (7 kpc, 170 km/s). In the $V_\phi - R$ by $\delta [M/H]_{R_G}$ plot (Panel M), there are two extended metallicity excess structures located near the bottom dashed black line. There is also a smaller faint metallicity excess substructure located in the bottom right quadrant however, it is less extended and has a lower metallicity residual amplitude compared to the other two structures. The spatial distribution of the metallicity variations in the bottom plot (Panel R) shows slight azimuthal metallicity variations. In this model, there is one major metallicity excess structure in the annulus between 4-6kpc. Along the 6kpc circle, it is shown that the residual metallicity varies from metal rich to metal poor with increasing azimuthal angle. There also exists another metallicity excess substructure at the outer edge of the simulation region.

In the fourth column from left, we show the results of the \texttt{LSB} model. In Panel I, there are less kinematic substructures present. The uppermost dot-dashed line tracks the only major kinematic substructure we identified. This ridge is not as extended as those seen in the previous simulations that included transient spiral arms. The inward-moving regions do not have extended prominent bands but still show evidence of some substructure in the form of patchy overdensities in the velocity space. On the other hand, the metallicity distribution in velocity space (Panel N) reveals many thin metallicity substructure bands, altering between metal-rich and metal-poor, across galactocentric radius. These bands are significant thinner than the bands that appear in the models with spiral arms. As with all other models, the spatial metallicity distribution (Panel S) also demonstrates the presence of azimuthal metallicity variation.

The final column displays the outcome of the \texttt{Sgr Multi-impact} model. In Panel J, we see that a satellite can create many kinematic ridges in the plane of the disk from tidally-induced spirals arms. Similarly, we see many $\delta[M/H]_{R_G}$ ridges in the velocity space (Panel O) that appear to be somewhat aligned with the kinematic substructure. However, the amplitude of the azimuthal metallicity variations in Panel T are significantly weaker compared to that of the azimuthal metallicity variations from the transient spiral arms and bar models.

In all of our simulations, we observe kinematic ridges across the $V_\phi$–$R$ plane and azimuthal metallicity variations indicating that spiral arms, the bar, and satellites can independently generate these substructures, as noted by previous studies (see introduction). It is also possible that a combination of these Galactic structures can produce these two types of disequilibrium features. However, a striking result emerges when coloring the stars in the $V_\phi$–$R$ plane by metallicity excess. In the models that include the perturbational effects of a bar and transient spiral arms there exists a strong correlation between the locations of the kinematic ridges and the metallicity substructure. This correlation is less-defined for our model that does not include the effects of transient spiral arms in the disk.

\section{Discussion}
\label{sec:disc}
We now turn to interpreting the results and highlight how  dynamical processes shape the observed metallicity variations in the Galactic disk. We found that the azimuthal metallicity variations are aligned with the kinematic ridges in the $V_{\phi} - R$ plane seen in the \textit{Gaia} data. The strong alignment between the azimuthal metallicity variations and the kinematic ridges observed in the \textit{Gaia} $V_{\phi} - R$ plane (see Figure \ref{fig:three_panel_ridges}) suggest that these two features could be related. If the mechanism responsible for shaping the stellar kinematic ridges and metallicity variations were independent, we may not expect such a close alignment between the two substructures. Our comparison with the simulations allows us to probe under what circumstances non-axisymmetric potentials can produce the alignment between the kinematic ridges and  metallicity variations. 

The results of our simulations have two major implications for the imprint of non-axisymmetric structures on the stellar chemistry and kinematic distribution of the MW. First, the \texttt{LSB} model demonstrates that while a long slow bar of constant pattern speed is capable of producing both kinematic substructure and azimuthal metallicity variations, it alone is insufficient to produce the alignment between the kinematic ridges and metallicity variations. In the 4th column from left of Figure \ref{fig:megaplot}, there are at least two prominent metal-rich substructures in Panel N that have no associated kinematic ridge in the velocity space of Panel I. There are also several weaker metal-rich substructures at lower $V_\phi$ with no direct mapping to any kinematic ridge. As a result, it appears that the transient spiral arms, provide the additional perturbations necessary to migrate stellar orbits on top of the bar-induced effects. Second, the models that most closely reproduce the chemo-kinematic alignment in the \textit{Gaia} data are the ones that include a bar and transient spiral arms. Both the \texttt{LSB+Spiral} and \texttt{SFB+Spiral} models produces multiple alternating inward and outward moving kinematic ridges, creates strong azimuthal metallicity variations, and shows a qualitative alignment between the kinematic ridges and metallicity variations. 

Of the two bar and transient spiral arm models, the \texttt{LSB+Spiral} model most closely reproduces the observed features and amplitudes in our Galaxy. The locations and number of metallicity excess substructures in the velocity space and the strength of the azimuthal metallicity variations closely resembles those seen in the \textit{Gaia} data. The parameters of the model are in agreement with the ``long-slow bar" model that is favored by recent observational constraints of the bar's length and pattern speed \citep{Wegg2015+, Clarke2022+, Dillamore2024+, Zhang2024+}

While the alignment of the kinematic and metallicity substructure is shown at one snapshot in Figure \ref{fig:megaplot}, the coherence between the two structures exists over multiple time steps. In Figure \ref{fig:TimeEvolution}, we show the time evolution of the $V_\phi - R$ plane colored by $\delta[M/H]_{R_G}$ for a series of time steps ranging from t=-417 Myr ago to t=0 Myr for the \texttt{LSB+Spiral} model. It is worth noting that \citet{Hunt2019+} identified the snapshot at t=-174 Myr ago as providing the best match to the observed kinematic structure. The contour lines overlaid on each panel are regions where $V_R = 0$ and trace out the boundaries of the outward and inward moving kinematic ridges. While the locations of the kinematic and metallicity substructures do shift from panel to panel, the correlation between the two remain fairly aligned throughout the range of time steps shown. The persistent of the chemo-kinematic alignment across time suggest that the correlation is not a transient phenomenon but rather an imprint of dynamical processes associated with the bar and spiral arms.

We also explored whether repeated interactions with a Sgr-like galaxy could produce similar substructure and chemo-dynamical alignment. Although our satellite model produces similar kinematic and metallicity substructure, it fails to produce azimuthal metallicity variations that are of comparable strength to those variations generated by the bar and spiral arm models. The maximum amplitude of the azimuthal metallicity variations due to our satellite model is about a factor of three weaker than the maximum amplitude of the variations due to the bar and spiral arm models. While the absolute amplitude will vary depending on the assumed initial radial metallicity gradient, the relative strength between models should not change by much.

Figure \ref{fig:ChangeInRg} shows the change in guiding center radius, $\Delta R_G = R_{G,f} - R_{G,i}$, for stars contained within a 3.5 kpc radius around the sun at the present-day snapshots. $R_{G,i}$ is the guiding center radius of the star at the start of the simulation and $R_{G,f}$ is the guiding center radius at the present-day snapshot. Stars that end up located further than their birth radius, called outwards migrators, have $\Delta R_G > 0$. Meanwhile, stars that migrated inwards have $\Delta R_G < 0$. There is a large fraction of outwards migrators seen in the \texttt{LSB+Spiral} and \texttt{SFB+Spiral} models. On the other hand, the shape of $\Delta R_G$ is centered more closely around zero for the satellite model. This helps to explain the lack of strong azimuthal metallicity variations from the satellite because the radial metallicity profile is defined as a function of the star's guiding center radius. In this approach, each value of metallicity excess directly corresponds to a change in the guiding center radius. A star with $\Delta R_G > 0$ is metal-rich relative to its predicted metallicity and a star with $\Delta R_G < 0$ is metal-poor relative to its predicted metallicity from the linear metallicity function. Taken together, this suggest that our model of a multi-impact satellite can not generate azimuthal metallicity variations of strength comparable to those generated by the bar and spiral arm models, limiting the parameter space of satellite models that can generate strong azimuthal metallicity variations. 

We also note that the results of our Sgr-like model differs from that of \citet{Carr2022+} who reported stronger azimuthal metallicity variations. This is unsurprising given the differences in the model setup, the calculation of the metallicity excess, and our choice of simulation snapshot to show. In particular, \citet{Carr2022+} showed that the azimuthal metallicity variations are the strongest immediately after recent Sgr passages but weaken as the disk phase-mixes. Additionally, \citet{Carr2022+} mentions that secular processes, such as the bar and spirals, are likely more important dynamical drivers of migration in the inner Galaxy. Further out, beyond the solar neighborhood, satellite bombardment should play a bigger role due to the longer mixing times and the weaker restoring potential in the outer Galaxy. Together, these results highlight how sensitive metallicity signatures are to the initial conditions of the merger, the dynamical history of the interaction, the initial metallicity distribution in the disk, and the definition of metallicity excess.

Although the \texttt{LSB+Spiral} model is our best-fitting qualitative match to the \textit{Gaia} data, it does not perfectly reproduce every feature seen in the \textit{Gaia} data. This is not surprising given the limitations of our test-particle simulations. Our simulations integrate the orbits of massless star particles in time-dependent analytic gravitational potentials and so does not include self-gravity, gas, star formation, or chemical evolution. As a result, there are some missing physics (e.g., gas-star interactions, feedback, etc.) that could be important for the presence of azimuthal metallicity variations and the alignment with the co-moving groups but are not captured. Additionally, the treatment of the satellite (see section~\ref{subsec:satellitesetup}) is a simplified model for Sgr with fixed mass and a last pericenter passage that occurred earlier than those from live N-body simulations \citep{Laporte2018+, Carr2022+}. We do not attempt to fit for unique Galactic parameters (e.g. bar pattern speed, bar pattern length, spiral arm shape, Sgr mass and dynamical history) or to assign a single mechanism to individual features of the chemo-kinematic substructure. Even with these caveats, there is value in using test-particle simulations as seen in works done by \citet{Antoja2014+, Hunt2018+, Hunt2019+, Gandhi2022+} which demonstrated that non-axisymmetric structures can create kinematic substructure in the disk. Follow-up work will include some range of these missing physics for a more complete picture.

Despite these limitations, our goal in this paper is to demonstrate that non-axisymmetric perturbations can qualitatively reproduce the observed chemo-kinematic alignment between the co-moving groups and metallicity variations and illustrate how different combinations of non-axisymmetric structures produce distinct patterns in the $V_\phi - R$ plane and $\delta [M/H]_{R_G}$ distribution. We are successful in determining that the bar and spiral arms produces a strong correlation between the kinematic ridges and metallicity variations. In addition, the strength of the azimuthal metallicity variations is maximized in models that have a bar and spiral arms.

The qualitative best-fitting model is the \texttt{LSB+Spiral} model which provides valuable clues by identifying the regions where the metallicity patterns do not match the kinematic ridges. In the second column, upper middle panel of Figure \ref{fig:megaplot}, the uppermost dashed black line traces the slope and approximate outer edge of an outward-moving velocity substructure visible in the left panel. While a corresponding metallicity excess substructure aligns with this ridge, it extends well beyond the width of the kinematic substructure and transitions into a metal-poor region at higher $V_{\phi}$. This behavior is not observed in the \textit{Gaia} data and may indicate the influence of additional dynamical processes not captured in our simulations. However, this is a relatively minor effect and the global alignment between the kinematic ridges and azimuthal metallicity variations suggest that the bar and spiral arms are one of the dominant mechanisms driving the correlation between the two.

\section{Conclusion}
\label{sec:conclude}
In this work, we explored the origins of azimuthal metallicity variations in the Galactic thin disk and their connection to the stellar kinematics using \textit{Gaia} DR3 and test-particle simulations of the MW. We began by recovering the negative radial metallicity profile of the thin disk and confirming the presence of azimuthal metallicity variations. We find an alignment between the azimuthal metallicity variations and the kinematic substructure of the \textit{Gaia} disk stars. To interpret the correlation between these two, we compared the data to a suite of MW simulations with varying combinations of bar and spiral arm patterns.  The main findings are summarized below:

\begin{itemize}

    \item {We measure a negative radial metallicity profile of $[M/H] = -0.057R_G + 0.41$ and $[M/H] = -0.062R + 0.45$ for our thin disk sample stars from \textit{Gaia} DR3. The results for the radial metallicity profiles are consistent with other studies \citep[e.g.][]{Friel2002+, Luck2011+, Hayden2014+,Onal2016+, Hawkins2023, Hackshaw2024, Akbaba2024}. Using our modeled 1-D metallicity profile, we explore the 2D metallicity distribution of the galaxy and search for azimuthal metallicity variations. We find evidence for azimuthal metallicity variations on the order of $\sim 0.1$ dex (Section \ref{subsec:results_rmg}), consistent with other studies \cite[e.g.][]{Poggio2022, Hawkins2023, Hackshaw2024}. }

     \item {Dynamical process due to the bar and spiral arms, such as resonances and perturbations, are main drivers in shaping the kinematic substructure and can therefore generate metallicity variations. From our simulations, these galactic structures will non-axisymetrically migrate stars away from their birth locations, where they inherited the metallicity from that location, and move these stars to new regions of the Galaxy, where their chemistry does not reflect the local chemistry.}
    
     \item {The comparison of the \textit{Gaia} data with the MW models suggest that while the Galactic bar can generate both kinematic substructure and azimuthal metallicity variations, it is insufficient to produce the observed alignment between the kinematic ridges and metallicity variations (Section \ref{subsec:comparesims}). }

     \item Our simulation of repeated interactions with a Sgr-like dwarf galaxy produces kinematic and metallicity substructure, but falls short at producing strong azimuthal metallicity variations compared to the bar/spiral models. The amount of radial migration driven by the satellite is significantly weaker compared to the bar + spiral arm models. This result, in combination with the findings from \citet{Carr2022+}, places constraints on the parameter space over which Sgr-like interactions can drive azimuthal metallicity variations seen in the Galaxy (Section \ref{sec:disc}). 

     \item {The \texttt{LSB+Spiral} model is the qualitative best-fit to the \textit{Gaia} data because it produces multiple ridges in the $V_\phi$–$R$ plane, creates strong azimuthal metallicity variations, and shows a qualitative alignment between these two substructures. The alignment between the chemo-kinematic substructure in our \texttt{LSB+Spiral} model (Figure \ref{fig:TimeEvolution}) persists across multiple timesteps, indicating that the correlation is not a transient feature. However, the model does not reproduce all of the detailed features seen in the \textit{Gaia} sample, suggesting that there may be other origins for at least some part of the observed azimuthal metallicity variations or that our model galaxy does not reflect the complete dynamical history of the MW.}
    
 \end{itemize}

Taken together, these findings support the interpretation that the azimuthal metallicity variations are not solely a product of stellar birth conditions but rather have been shaped by dynamical processes associated with the Galactic bar and spiral arms of the MW. Further work incorporating more detailed models and additional observational constraints will be crucial for refining our understanding of how the MW dynamics shapes the metallicity distribution. 

\section*{Acknowledgements}
We thank the anonymous referee for constructive feedback
on the manuscript. CJ thanks Dionysis Gakis for useful discussions that helped improve this work. KH is partially supported by NSF AST-2407975. KH acknowledge support from the Wootton Center for Astrophysical Plasma Properties, a U.S. Department of Energy NNSA Stewardship Science Academic Alliance Center of Excellence supported under award numbers DE-NA0003843 and DE-NA0004149, from the United States Department of Energy under grant DE-SC0010623. This work was performed in part at the Aspen Center for Physics, which is supported by National Science Foundation grant PHY-2210452.

We make use of data from the European Space Agency (ESA) mission \textit{Gaia} (\url{http://www.cosmos.esa.int/gaia}), processed
by the \textit{Gaia} Data Processing and Analysis Consortium (DPAC;
\url{http://www.cosmos.esa.int/web/gaia/dpac/consortium}). Funding for the DPAC has been provided by national institutions, in particular the institutions participating in the \textit{Gaia} Multilateral Agreement. JH acknowledges the support of a UKRI Ernest Rutherford Fellowship ST/Z510245/1.

\bibliography{references}

\appendix 

\section{Sensitivity of Azimuthal Metallicity Variations On R or $R_G$}
\label{appendix: R vs R_G}
Throughout this work, we defined metallicity excess as a function of the star's guiding center radius. This choice was motivated by the fact that $R_G$ better reflects a star's average orbital radius and helps to minimize the effects of epicyclic blurring. Because stars execute radial epicycles about their guiding center radii, using the instantaneous R mixes stars with different $R_G$ at different epicyclic phases. This artificially smears out the coherence of the metallicity variations in velocity space. However, since many studies in the literature have defined metallicity profiles and metallicity excess using the star's galactocentric radius \citep[e.g.][]{Hayden2014+, GaiaDR3Chemistry, Hawkins2023, Hackshaw2024}, we include Figure \ref{fig:three_panel_galactocentric}. These panels demonstrate that azimuthal metallicity variations remain visible when defining metallicity excess as a function of R and confirm that our overall results are not artifacts of our choice of coordinates. 

In the middle panel of Figure \ref{fig:R_ridges_and_variations}, we show the $V_\phi$–$R$ plane colored by metallicity excess as a function of $R$. While the alignment between the metallicity excess substructure and kinematic ridges is still apparent there is a metallicity gradient that appears along individual ridges. Along a given ridge, there is a transition from metal-poor (blue) regions at high $V_\phi$ to metal-rich (red) at low $V_\phi$. This is the result of stars at a given radius but with different epicyclic phases. It's been shown that some of the kinematic ridges are roughly along lines of constant angular momentum \citep{Martinez-Medina2019+, Fragkoudi2019+} and so stars have nearly the same guiding center radius. Thus, for stars at some guiding center radius with an azimuthal angle $\theta_R = 0$ (pericenter), it will appear to be metal-poor relative to the stars in the immediate neighborhood. Meanwhile, when that star is at an  azimuthal angle $\theta_R = \pi$ (apocenter), it will appear metal-rich relative to the stars around it. 

\begin{figure}[h]
    \centering
    \includegraphics[width=1\linewidth]{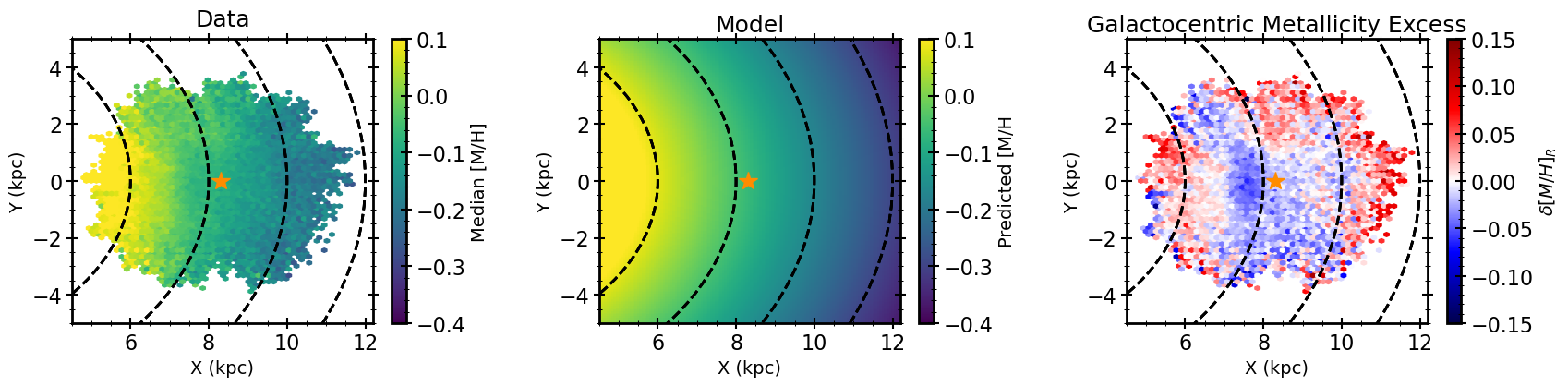}
    \caption{Same as Figure \ref{fig:three_panel} but now the radial metallicity profile is calculated as a function of galactocentric radius. In this case, we denote metallicity excess as $\delta [M/H]_{R}$ to distinguish from the metallicity excess calculated with a guiding center radius metallicity profile ($\delta [M/H]_{R_G}$). }
    \label{fig:three_panel_galactocentric}
\end{figure}

\begin{figure}
    \centering
    \includegraphics[width=1\linewidth]{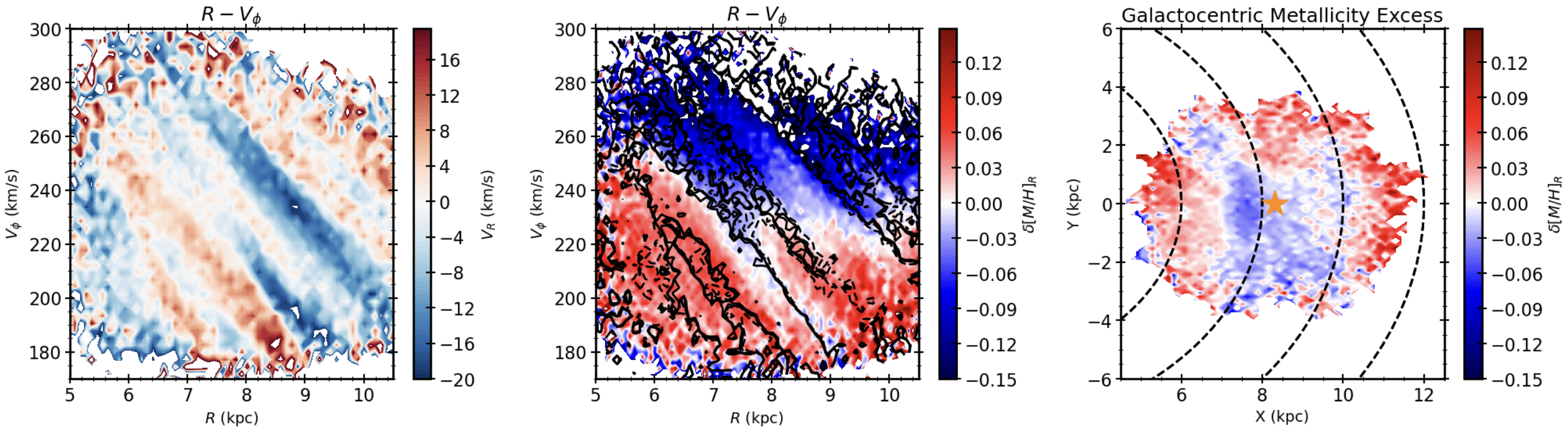}
    \caption{ \textbf{Left Panel:} Distribution of azimuthal velocity ($V_{\phi}$) as a function of galactocentric radius (R), colored by $V_{R}$. \textbf{Middle Panel:} Same as left panel but colored by $\delta [M/H]_{R}$. The contours in each panel are regions where $V_R=0$ and are used to trace the locations of the kinematic ridges in the left panel.  \textbf{Right Panel:} $\delta [M/H]_{R}$ in the X-Y plane.}
    \label{fig:R_ridges_and_variations}
\end{figure}

\end{document}